\documentclass[bold,12pt]{article}
\usepackage[nocompress]{cite}
\usepackage{amsfonts}
\usepackage{graphicx}
\usepackage{multirow}
\usepackage{geometry}
\usepackage{subfigure}
\newtheorem{theorem}{Theorem}

\begin{document}

\begin{titlepage}

\title{Modularity functions maximization with nonnegative relaxation facilitates community detection in networks}
\author{Jonathan Q. Jiang${}^{1,*}$,~~Lisa J. McQuay${}^2$\\[0.2cm]
\small $^1$Department of Computer Science, City University of Hong Kong,\\
\small 83 Tat Chee Avenue, Kowloon, Hong Kong\\
\small Email: qiajiang@cityu.edu.hk\\[0.1cm]
\small $^2$ RTI Health Solutions, Research Triangle Park, NC, USA\\
\small Email: ljmcquay@rti.org\\[0.2cm]
\small $*$ Corresponding author. Tel: +852-219 420 26; Fax: +852-219 420 26}

\date{}
\maketitle

\noindent\rule[0.2\baselineskip]{\linewidth}{0.5pt}\\[0.2cm]
\textbf{Abstract}\\
We show here that the problem of maximizing a family of quantitative functions, encompassing both the modularity ($Q$-measure) and modularity density ($D$-measure), for community detection can be uniformly understood as a combinatoric optimization involving the trace of a matrix called {\em modularity Laplacian}. Instead of using traditional spectral relaxation, we apply additional nonnegative constraint into this graph clustering problem and design efficient algorithms to optimize the new objective. With the explicit
nonnegative constraint, our solutions are very close to the ideal
community indicator matrix and can directly assign nodes into communities. The near-orthogonal columns of the solution can be reformulated as the posterior probability of corresponding node belonging to each community. Therefore, the proposed method can be exploited to identify the fuzzy or overlapping communities and thus facilitates the understanding of the intrinsic structure of networks. Experimental results show that our new algorithm consistently, sometimes significantly, outperforms the traditional spectral relaxation approaches.\\[0.2cm]
\noindent\rule[0.2\baselineskip]{\linewidth}{0.5pt}

\end{titlepage}

\section{Introduction} \label{sec1}

Networks have become a popular tool for describing complex real-world systems. Typical examples include the Internet, metabolic networks, food webs, neural networks, technological networks, communication and distribution networks, and social networks (see \cite{Bar02} and reference therein for more examples). It has been observed that methods for associating a suitable community structure to a given network can support our analysis. Adopting a rather coarse-grained top-down point of view, one can elucidate a network's intrinsic structure and reveal its global or overall organization in terms of its \textit{modules} or \textit{communities}, i.e., suitably chosen collections of disjoint groups of nodes such that the nodes within each group are joined together in a tightly-knit fashion, while only looser connections exist between the nodes in different groups \cite{New01,Gir02}.

This description is just an intuitive concept rather than a rigorous definition, and thus is far way from meeting the requirement for detecting communities in graphs via computational algorithms. A first quantitative measure for evaluating the \lq \lq goodness of fit\rq \rq of a partition to a graph, the {\em modularity function} $Q$, was proposed by Newman and Girvan in \cite{New04}. Using this or similar such functions, community detection gets reduced to optimization problems, which has led to the emergence of a considerable number of community-detection procedures (see \cite{Por09,For10} for recent reviews). The modularity function $Q$, however, has been shown to have limited resolution, i.e., communities smaller than a certain scale may not be revealed by optimization of $Q$ even in the extreme case that they are cliques connected by single bridges\cite{For07}. To address this issue, another measure, namely the {\em modularity-density} or $D$-measure \cite{Li08} was recently proposed. As suggested by theoretical analysis as well as by numerical tests, optimizing this new measure does not seem to exhibit the resolution limit of the $Q$-measure.

Unfortunately, these two optimization problems are both NP-hard \cite{New04,Li08}. Inspired by the idea of spectral graph clustering, White and Smith \cite{Whi05} showed that maximizing the (original) $Q$-function can be reformulated as a spectral relaxation problem, i.e., an eigenvector problem involving a matrix called the {\em $Q$-Laplacian}, implying that modularity-based clustering can be understood as a special instance of spectral clustering---see also \cite{Jia09} where it was shown that this spectral-clustering approach can be extended quite naturally to also taking account of edge-weight data. The main disadvantage of such spectral relaxation approach is that the eigenvectors have mixed-sign entries, which could severely deviate from the true clustering indicator vectors (that these eigenvectors approximate). Therefore, we often require resorting to other clustering methods, such as K-means, to obtain final cluster results\cite{Whi05,Jia09}.

To go beyond previous works, we propose here a new method to optimize the modularity functions with additional nonnegative constraint. With the explicit nonnegative constraint, the solutions are very close to the ideal community indicator matrix and can be directly used to assign cluster labels to nodes. We give efficient algorithms to solve this problem with the nonnegative constraint rigorously. Experimental results show that our algorithm always converges and the performance is significantly improved in comparison with the traditional spectral relaxation approach.

The rest of this paper is organized as follows. Section \ref{sec2} shows that the problem of maximizing a family of modularity functions, encompassing both $Q$-measure and $D$-measure, for community detection can be reformulated as a combinatoric optimization involving the trace of a matrix called {\em modularity Laplacian}. Our proposed nonnegative relaxation approaches are introduced in Section \ref{sec3}. We emphasize the soft clustering nature of this method, which facilitates identifying fuzzy communities of networks in Section \ref{sec4}. Experimental results on synthetic as well as real-world networks are reported in Section \ref{sec5}. Finally, we conclude our work in Section \ref{sec6}.

\section{Problem reformulation} \label{sec2}

\subsection{Modularity functions}\label{sec2:sub1}

Our start point is a family of quantitative modularity functions. Let $G = (V,E,W)$ be a network with $n$ nodes,
where $V$ is the set of nodes, $W = (w(u,v))_{(u,v) \in E}$ is the weight matrix and $w(u,v)$ is the weight for
the edge connecting the nodes $u$ and $v$. For any two subsets $V_1, V_2$ of $V$, let $L(V_1,V_2): = \sum_{u \in V_1, v \in V_2} w(u,v)$ denote the sum over the weights of all edges connecting a vertex $u \in V_1$ and a vertex $v \in V_2$. Given a partition $\Pi$ of the network $G$, the modularity function $Q$-measure is defined as \cite{New04}
\begin{equation}
Q(\Pi): = \sum_{k=1}^K \left[\frac{L(V_k,V_k)}{L(V,V)} - \left(\frac{L(V_k,V)}{L(V,V)}\right)^2\right]\label{eq1}
\end{equation}
This measure provides a means to determine whether or not a partition is good enough to decipher the community structures of a network. Generally, a larger $Q$ value corresponds to a better community structure identification. As reported in \cite{For07}, $Q$-measure has been exposed
to resolution limit since the size of a detected module depends on the size of the whole network. Recently, Li and his collaborators \cite {Li08} proposed a quantitative measure called \textit{modularity density} D-measure to resolve this difficulty. For a specific network partition $\Pi$, the modularity density $D(\Pi)$ is given by the sum
\begin{equation}
D(\Pi): = \sum_{k=1}^K \frac{L(V_k,V_k) - L(V_k, \overline{V}_k)}{|V_k|}\label{eq2}
\end{equation}
Obviously, it can be referred to as a combination
of the ratio association (the first term) and the ratio cut (the second term) and therefore be naturally extended to a family of quantitative functions
\begin{equation}
D_\lambda(\Pi): = \sum_{k=1}^K \frac{(1-\lambda)L(V_k,V_k) - \lambda L(V_k, \overline{V}_k)}{|V_k|} \label{eq3}
\end{equation}
where $0 \le \lambda \le 1$ is the trade-off parameter. When $\lambda = 1$, $D_\lambda$ is the ratio cut;
when $\lambda=0$, $D_\lambda$ is the ratio association; when $\lambda=0.5$, $D_\lambda$ is the modularity density $D$.

Let $\mathbf{x}_k, k = 1, \dots, K$ be the community indicators where the $u$-th element of $\mathbf{x}_k$ is $1$ if the node $u$ belongs to community $k$, and $0$ otherwise. For example, if nodes within each community are adjacent, then $\mathbf{x}_k = (0, \dots, 0, \overbrace{1, \dots, 1}^{n_k}, 0, \dots, 0)$. If we further define $Q = (\mathbf{q}_1, \dots, \mathbf{q}_K)$ as $\mathbf{q}_k = \frac{\mathbf{x}_k}{\|\mathbf{x}_k\|}$, the previous works \cite{Jia09,Whi05} show that maximization of modularity function can be reformulated as follows
\begin{equation}
\max_Q \mathrm{Tr} Q^T (\mathcal{W} - \mathcal{D}) Q~~~~~\mathrm{s.t.}~~Q^TQ = I, Q \ge 0 \label{eq4}
\end{equation}
where $\mathcal{W}: = \frac{W}{\mathrm{vol}G}$, $\mathcal{D}:=\frac{\mathbf{d}^T\mathbf{d}}{\mathrm{vol}G^2}$, $\mathbf{d} = (d_1, \dots, d_n)^T$ such that component $d_u$ equals the weighted degree of node $u$ and $\mathrm{vol}G = L(V,V)$ is the total number of edges in the network. Clearly, this is a combinatoric optimization, i.e., a matrix trace maximization with nonnegative and orthogonal constraints. We will demonstrate here that maximizing the family of the quantitative functions (\ref{eq3}) can be represented as the same type of problems.

We can easily see that $L(V_k, \overline{V}_k) = \sum_{u \in V_k, v \in \bar{V}_k} w(u,v) = \mathbf{x}_k^T (D - W)\mathbf{x}_k$ and $L(V_k, V_k) = \mathbf{x}_k W \mathbf{x}_k$ where $D$ is the degree matrix of network $G$. Thus, the objective function can be rewritten as
\begin{eqnarray}
D_\lambda & = & \sum_{k=1}^K \frac{(1-\lambda) \mathbf{x}_k^T W \mathbf{x}_k - \lambda \mathbf{x}_k^T(D - W)\mathbf{x}_k}{\mathbf{x}_k^T \mathbf{x}_k}\nonumber\\
& = & \sum_{k=1}^K \frac{\mathbf{x}_k^T(W - \lambda D)\mathbf{x}_k}{\mathbf{x}^T_k \mathbf{x}_k} \label{eq5}
\end{eqnarray}
which yields
\begin{equation}
\max_Q D_\lambda = \mathrm{Tr} Q^T (W - \lambda D)Q~~~~~\mathrm{s.t.}~~Q^TQ = I, Q \ge 0 \label{eq6}
\end{equation}
Taken together, Eqs (\ref{eq4}) and (\ref{eq6}) are the same type of optimization problem, i.e., a matrix trace maximization with nonnegative and orthogonal constraints which can uniformly represented as
\begin{equation}
J_\mathrm{mmn}= \max_Q \mathrm{Tr} Q^T (-M)Q~~~~~\mathrm{s.t.}~~Q^TQ = I, Q \ge 0 \label{eq7}
\end{equation}
The matrix $M$, corresponding to the modularity function used for investigation, plays a similar role to the combinatoric Laplacain matrix in graph theory \cite{Whi05,Jia09}. Analogously, we call it \textit{modularity Laplacian} hereafter.

\subsection{Spectral embedding and clustering} \label{sec2:sub2}

Note that there is a canonical one-to-one correspondence between subsets $U \subset V$ and $\{0,1\}$-vectors $\mathbf{x} \in \Re^n$ given by associating to each such vector $\mathbf{x} \in \Re^n$ its \textit{support} $\mathrm{supp}(\mathbf{x}):=\{u \in V: \mathbf{x}(u) \ne 0\}$ that induces between a canonical one-to-one correspondence between partitions $\Pi$ of $V$ and collections $X = \{\mathbf{x}_1, \dots, \mathbf{x}_K\}$ that sum up to the "all-one vector" $\mathbf{1}_n \in \Re^n$, i.e., the map $\mathbf{1}_n: V \rightarrow \Re$ that maps every $v \in V$ onto $1$. In consequence, the measure of a partition $\Pi$ of $V$ of rank $K$ is bounded from (\ref{eq7}) by the supremum $\mathrm{sup}(\sum_{\mathbf{e}\in \mathcal{E}} \mathbf{e}^T (-M)\mathbf{e}: \mathcal{E})$ where $\mathcal{E}$ runs over all collections of $K$ pairwise orthogonal unit vectors in $\Re^n$ and, hence (in view of the fact that this supremum coincides with the sum of the $K$ smallest eigenvalues of the symmetric matrix $M$ --- see e.g., Section II in \cite{Jia09}), $J_\mathrm{mmn} \ge \sum_{i=1}^K \lambda_i$ where $\lambda_1 \le \lambda_2 \le \dots \le \lambda_n$ is the sequence, in ascending order, of the necessarily real eigenvalues of the symmetric matrix $M$.

Thus, following the ideas presented in \cite{Whi05,Jia09}, we can construct reasonably good community structures as follows. For each $k = 1, \dots, n$, we may choose an eigenvector $\mathbf{e}_k$ of $M$ with eigenvalue $\lambda_k$  so that the $\mathbf{e}_k (k=1,\dots,n)$ form an orthonormal family of eigenvectors of $M$ and then consider, for each $K=1,2,\dots,n$ and
the associated {\em spectral embedding}
$\chi_K$ of $V$ into $\Re^{\{1,2, \dots,K\}}$ that maps any $v\in V$ onto the point
$$
\chi_K(v):\{1,2, \dots,K\}\rightarrow \Re: k\mapsto \mathbf{e}_k(v)
$$
in $\Re^{\{1,2, \dots,K\}}$. If the $\mathbf{e}_k$ were of the form $\mathbf{e}_k=\mathbf{q}_k$ for some collection $\mathbf{x}_1,\dots,\mathbf{x}_K$ of $K$ (necessarily pairwise orthogonal) $\{0,1\}$-vectors in $\Re^n$ with $\sum_{k=1}^K \mathbf{x}_k=\mathbf{1}_n$
associated to a $K$-partition $\Pi$, the support of any vector of the form $\chi_K(v)$ for some $v\in V$ would be the one-element subset $\{k\}$ with $v\in\mathrm{supp} (\mathbf{x}_k)=\mathrm{supp}(\mathbf{e}_k)$, and two vectors of the form $\chi_K(u)$ and $\chi_K(v) \,\,(u,v \in V)$ would coincide if and only if $u$ and $v$ would belong to the same subset $U$ in $\Pi$.

\section{Modularity function maximization with nonnegative relaxation} \label{sec3}

From its definition,  only one element is positive and others are zeros in each row of indicator matrix $Q$. Thus, we require a solution which optimizes a quadratic function of $Q$ with two constraints: (1) orthonormal and (2) nonnegative. These difficult constraints should be relaxed to make the problem trackable. If we retain orthogonality while ignore the nonnegativity, the solution is the aforementioned spectral relaxation approach. The main disadvantage of such relaxation is that the eigenvectors have mixed-sign entries which may severely deviate from the true community indicator vectors. Therefore, most previous applications resort to a two-step procedure \cite{Whi05,Jia09}: (1) embedding the network into the eigenvector space and (2) clustering these embedded points using other algorithm, such as K-means clustering.

A more accurate relaxation is adding the nonnegative constraints on $Q$. One can see that when orthogonal and nonnegative constraints are satisfied simultaneously, the solution will very close to the ideal indicator matrix and thus can be used directly to assign community labels to nodes. This motivates our nonnegative relaxation approach for the modularity function optimization. Formally, the optimization of Eq (\ref{eq7}) is mathematically identical to
\begin{equation}
\max_Q \mathrm{Tr}Q^T(\rho I - M) Q~~~~~\mathrm{s.t.}~~Q^TQ = I, Q \ge 0 \label{eq8}
\end{equation}
because the $\rho$ term $\mathrm{Tr}Q^T \rho I Q = \rho \mathrm{Tr}I = \rho n$ is a constant. In particular, we set $\rho = \lambda_\mathrm{max}(M)$ to be the largest eigenvalue of $M$ such that $\rho I - M$ is positive definite. This transformation makes the optimization a well-behaved problem. We begin with the Lagrangian function
\begin{equation}
\mathcal{L} = J_\mathrm{mmn}(Q) -\mathrm{Tr}\Lambda(Q^TQ - I) - \mathrm{Tr}\Sigma Q \label{eq9}
\end{equation}
where the Lagrange multiplier $\Lambda$ enforces the orthogonality condition $Q^TQ = I$ and the Lagrange multiplier $\Sigma$ enforces the nonnegativity of $Q \ge 0$. The KKT complementary slackness condition becomes \begin{equation}
\left(\frac{\partial \mathcal{L}}{\partial Q_{ik}}\right) Q_{ik} = [(\rho I - M)Q - Q \Lambda]_{ik} Q_{ik} = 0 \label{eq10}
\end{equation}
Clearly, a fixed point also satisfies
\begin{equation}
[(\rho I - M)Q - Q \Lambda]_{ik} Q^2_{ik} = 0
\label{eq11}
\end{equation}
which is mathematically identical to Eq(\ref{eq10}). Summing over $k$, we obtain $\Lambda_{ii} = [Q^T(\rho I - M)Q]_{ii}$. This gives the diagonal elements of $\Lambda$. To find the off-diagonal elements of $\Lambda$, we temporarily ignore the nonnegativity requirement and set $\partial L/\partial Q = 0$ which leads to $\Lambda_{ii'} = [Q^T (\rho I - M)Q]_{ii'}$. Combining these two results yields
\begin{equation}
\Lambda = Q^T (\rho I - M) Q \label{eq12}
\end{equation}
Decomposing $M$, $\Lambda$ into positive part and negative part as
$$
M = M^+ - M^- \quad \Lambda = \Lambda^+ - \Lambda^-
$$
where $M^+ = (|M| + M)/2$, $\Lambda^+ = (|\Lambda| + \Lambda)/2$, $M^- = (|M| - M)/2$ and $\Lambda^- = (|\Lambda| - \Lambda)/2$, respectively. Now concentrating on the variable $Q$ in $\mathcal{L}$, we have
\begin{eqnarray}
\frac{1}{2}\frac{\partial (J_\mathrm{mmn} - \mathrm{Tr} \Lambda Q^TQ)}{\partial Q}{}&=&{} \rho Q - M^+Q + M^-Q - Q\Lambda^+ + Q\Lambda^- \nonumber\\
{}&=&{}(\rho Q + M^-Q + Q\Lambda^-) - (M^+Q + Q\Lambda^+) \label{eq13}
\end{eqnarray}
As in Nonnegative Matrix Factorization (NMF) \cite{Lee01}, Eq (\ref{eq13}) leads to the following multiplicative update formula:
\begin{equation}
Q_{ik} \leftarrow Q_{ik} \sqrt{\frac{(\rho Q + M^-Q + Q \Lambda^-)_{ik}}{(M^+ Q + Q \Lambda^+)_{ik}}}
\label{eq14}
\end{equation}

We can see that using this update, $Q_{ik}$ will increase when the corresponding
element of the gradient in Eq (\ref{eq13}) is larger than zero, and will decrease otherwise.
Therefore, the update direction is consistent to the update direction in
the gradient ascent method. Note that the feasible domain of Eq (\ref{eq7}) is non-convex, indicating that our algorithm can only reach local optimizations. However, we show by empirical study (in Section \ref{sec5:sub1}) that our algorithm yields much better results than spectral clustering. In typical implementation of modularity function maximization with nonnegative relaxation algorithm, the computational complexity is $O(n^2K)$, which is similar to traditional spectral clustering approach. The correctness and convergence of the proposed algorithm are assumed by the following two theorems

\begin{theorem}
Fixed points of Eq (\ref{eq14}) satisfy the KKT condition of the optimization problem of Eq (\ref{eq7}). \label{the1}
\end{theorem}
\begin{theorem}
Under the update rule of Eq(\ref{eq13}), the Lagrangian function
\begin{equation}
L = {\rm Tr}[Q^T(\rho I - M )Q - \Lambda(Q^TQ - I)]
\label{eq15}
\end{equation}
increases monotonically.
\label{the2}
\end{theorem}
The proof of Theorem \ref{the1} could be straightforwardly implied in the derivation of the update rule (\ref{eq13}) and the proof of Theorem \ref{the2} is given in Appendix \ref{app}. As mentioned before, the solution is very close to the ideal class indicator
matrix due to the orthonormal and nonnegative constraints. Thus $Q$ can be directly used to assign cluster labels to data points. Specifically, the node $u$ is assigned to community $k^*$ as
\begin{equation}
k^* = \arg \max_k Q_{uk} \label{eq19}
\end{equation}

\section{Soft clustering and fuzzy community detection}\label{sec4}

In fact, the rows of indicator $Q$ are not exactly orthogonal since the off-diagonal elements of the Lagrangian multiplier are obtained by temporally ignoring the nonnegativity constraint. This slight deviation from rigourous orthogonality produces a benefit of soft clustering. An exact orthogonality implies that each row of $Q$ can have only one nonzero element, which indicates that each node belongs to only one community. This is \textit{hard clustering}, such as in K-means. The near-orthogonality condition relaxes this a bit, i.e., each node could belong fractionally to more than one community. This is \textit{soft clustering}.

An interesting phenomenon in community detection is that several nodes may not belong to a single community. Instead, assigning them into more than one group is more reasonable \cite{Rei04,Pal05}. Such nodes may mean a fuzzy categorization and thus take a special role, for example, signal transduction in biological networks. Another issue is that some nodes, considered as \textit{unstable} vertices in \cite{Gfe05}, locate on the border between two communities are hard to classify into any group. Therefore, uncovering these nodes and overlapping communities can facilitate understanding the intrinsic structure of networks.

We provide here a systematic analysis of the soft clustering of nodes based on the proposed algorithm. In previous related studies \cite{Ding05,Luo09}, the rows of $Q$ have been interpreted as posterior probability for graph partition. In other words, the magnitude of $Q_{uk}$ quantifies the degree that the node $u$ belongs to community $k$ (see Section \ref{sec5:sub2} for illustrative instances). Specifically, after getting $Q$, we first row-normalize it to $\sum_k Q_{uk} = 1$ and then infer how strong the nodes belong to the community according to their community entropies to reveal the so-called unstable vertices. The community entropy of node $u$ is defined as \cite{Jia09}
\begin{equation}
\xi_u : = - \sum_k Q_{uk} \log_K Q_{uk} \label{eq18}
\end{equation}
Obviously, the nodes with large entropy must be less stable.

\section{Numerical results}\label{sec5}

In this section, we extensively test the proposed algorithm on two artificial benchmark networks with a known community structure, including the ad hoc network with 128 nodes and the LFR benchmarks. After that, the algorithm is applied to three famous real-world networks: Zachary's karate-club network, journal citation network and American college football team network.

\subsection{Test on artificial networks\label{sec5:sub1}}

\subsubsection{Ad hoc benchmark netowrk\label{sec5:sub1:sub1}}

The ad hoc network has 128 nodes, which are divided into 4 communities with 32 nodes each. Edges are placed randomly with two fixed expectation values so as to keep the average degree of a node to be 16 and the average edge connections $z_{\rm out}$ of each node to nodes of other modules. This experiment was designed by Girvan and Newman \cite{Gir02} and has been broadly used to test community-detection algorithms \cite{Gir02,New04,Gfe05}.

\begin{figure}
\centering
\includegraphics[width=0.9\textwidth]{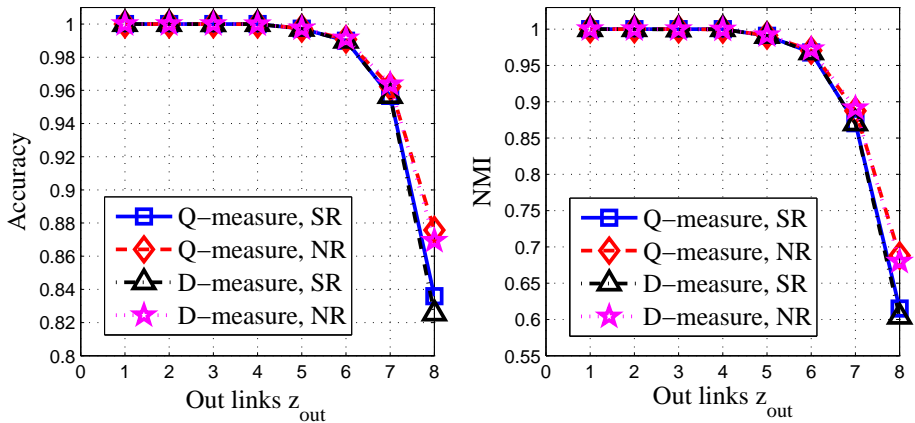}
\caption{The evolution of two metrics with respect to $z_\mathrm{out}$ for testing two methods on the ad hoc networks with known community structures. Each point is an average over 100 realizations of the networks. The term SR denotes the spectral relaxation approach and the term NR denotes the nonnegative relaxation approach, respectively.}
\label{fig1}
\end{figure}

We use the following two standard evaluation metrics to evaluate the performance for our method and spectral relaxation approach.$\\\\$
{\bf Accuracy} is calculated by:
\begin{equation}
\mathrm{Acc}:= \frac{\sum_{i=1}^n \delta(l_i, \mathrm{map}(c_i))}{n} \label{eq20}
\end{equation}
where $l_i$ is the true cluster label and $c_i$ is the obtained cluster label of $i$-th node, $\delta(x,y)$ is the delta function, and $map(\cdot)$ is the best mapping function. Note $\delta(x,y)=1$, if $x=y$; $\delta(x,y)=0$, otherwise. The mapping function $\mathrm{map}(\cdot)$ matches the true
class label and the obtained cluster label and the best mapping is solved by
Kuhn-Munkres algorithm \cite{Lov86}. A larger accuracy value indicates a better performance.$\\\\$
{\bf Normalized mutual information}(NMI): Given the sets of real communities $\mathbb{S}_r$ and found communities $\mathbb{S}_f$, the mutual information metric ${\rm MI}(\mathbb{S}_r,\mathbb{S}_f)$ is defined as
\begin{equation}
{\rm MI}(\mathbb{S}_r,\mathbb{S}_f) = \sum_{u \in \mathbb{S}_r, v \in \mathbb{S}_f} p(u,v) \log_2 \frac{p(u,v)}{p(u)p(v)}
\label{eq21}
\end{equation}
where $p(u), p(v)$ denote the probabilities that a node arbitrarily selected from the network belongs to the real community $\mathbb{S}_r$ and found community $\mathbb{S}_f$, respectively, and $p(u,v)$ denotes the joint probability that this arbitrarily selected node belongs to the communities $\mathbb{S}_r$ and $\mathbb{S}_f$ at the same time. The ${\rm NMI}(\mathbb{S}_r,\mathbb{S}_f)$ is defined as
\begin{equation}
{\rm NMI}(\mathbb{S}_r,\mathbb{S}_f) = \frac{{\rm MI}(\mathbb{S}_r,\mathbb{S}_f)}{\max(H(\mathbb{S}_r),H(\mathbb{S}_f))}
\label{eq22}
\end{equation}
where $H(\mathbb{S}_r)$ and $H(\mathbb{S}_f)$ are the entropies of $\mathbb{S}_r$ and $\mathbb{S}_f$ respectively.

\begin{table}
\caption{\label{tab1} Simulation results of our algorithm and spectral relaxation approach based on the general $\lambda$-measure when $z_{\rm out} = 5,6,7,8$ respectively. Results are averaged by 100 random realizations.}
\centering
\scalebox{0.8}{
\begin{tabular}{@{}ccccccc@{}}
\hline
\hline
\multirow{2}{*}{} & \multirow{2}{*}{} & \multirow{2}{*}{$\lambda$} & \multicolumn{4}{c}{$z_\mathrm{out}$}\\
\cline{4-7}
&  &  & 5 & 6 & 7 & 8\\
\hline
\multirow{12}{*}{Accuracy (\%)} & \multirow{6}{*}{SR} & 0 & 99.77$\pm$0.39&98.78$\pm$0.91&95.50$\pm$1.87&83.38$\pm$5.83\\
& & 0.2& 99.82$\pm$0.35&98.84$\pm$0.93&95.88$\pm$1.69&84.50$\pm$4.93\\
& & 0.4& 99.87$\pm$0.31&98.97$\pm$0.86&95.96$\pm$1.48&84.58$\pm$4.78\\
& & 0.6& 99.86$\pm$0.36&98.95$\pm$0.90&95.52$\pm$1.78&81.38$\pm$5.24\\
& & 0.8& 99.81$\pm$0.44&98.50$\pm$2.70&91.27$\pm$6.65&70.39$\pm$8.94\\
& & 1.0& 99.30$\pm$1.60&95.66$\pm$5.71&79.82$\pm$11.70&58.63$\pm$9.24\\
\cline{2-7}
& \multirow{6}{*}{NR} & 0 &99.81$\pm$0.35&98.97$\pm$0.84&96.38$\pm$1.58&87.30$\pm$6.03\\
& & 0.2&99.84$\pm$0.33&99.01$\pm$0.84&96.67$\pm$1.48&88.97$\pm$3.12\\
& & 0.4&99.86$\pm$0.32&99.06$\pm$0.88&96.66$\pm$1.47&88.77$\pm$3.93\\
& & 0.6&99.86$\pm$0.32&99.06$\pm$0.87&96.45$\pm$1.55&86.66$\pm$4.38\\
& & 0.8&99.85$\pm$0.33&98.78$\pm$2.25&92.22$\pm$8.73&71.14$\pm$11.32\\
& & 1.0&99.42$\pm$1.40&96.07$\pm$5.01&80.66$\pm$11.34&60.77$\pm$10.80\\
\hline
\multirow{12}{*}{NMI (\%)} & \multirow{6}{*}{SR} & 0& 99.24$\pm$1.26&96.18$\pm$2.75&86.67$\pm$5.08&60.75$\pm$7.58\\
& & 0.2&99.42$\pm$1.12&96.37$\pm$2.85&87.68$\pm$4.75&62.29$\pm$7.37\\
& & 0.4&99.57$\pm$1.01&96.73$\pm$2.70&87.86$\pm$4.13&62.62$\pm$6.76\\
& & 0.6&99.55$\pm$1.15&96.68$\pm$2.79&86.81$\pm$4.62&57.38$\pm$6.79\\
& & 0.8&99.39$\pm$1.43&95.80$\pm$4.86&79.07$\pm$9.82&44.48$\pm$8.98\\
& & 1.0&98.05$\pm$3.72&89.85$\pm$9.60&62.40$\pm$13.20&30.89$\pm$9.18\\
\cline{2-7}
&\multirow{6}{*}{NR} & 0& 99.39$\pm$1.14&96.74$\pm$2.63&89.04$\pm$4.45&68.44$\pm$7.70\\
& & 0.2&99.50$\pm$1.07&96.86$\pm$2.63&89.85$\pm$4.24&70.75$\pm$6.14\\
& & 0.4&99.55$\pm$1.03&97.03$\pm$2.78&89.85$\pm$4.23&70.56$\pm$7.31\\
& & 0.6&99.55$\pm$1.03&97.00$\pm$2.77&89.34$\pm$4.32&66.93$\pm$7.75\\
& & 0.8&99.52$\pm$1.05&96.55$\pm$3.54&83.51$\pm$10.64&48.85$\pm$10.46\\
& & 1.0&98.36$\pm$5.02&90.34$\pm$12.42&63.83$\pm$17.81&36.45$\pm$11.57\\
\hline
\hline
\end{tabular}}
\end{table}

Fig.\ref{fig1} shows the evolution of $Q$-measure and $D$-measure with respect to $z_{\rm out}$ by our method (NR) and spectral relaxation approach (SR), respectively. From the results, we can observe whatever modularity function and evaluation metric was utilized, our nonnegative relaxation method outperforms traditional spectral relaxation approach on all benchmark networks, especially for the most difficult case when $z_{\rm out} = 8$. Table \ref{tab1} give the simulation results of these two algorithms based on the general $\lambda$-measure when $z_{\rm out} = 5,6,7,8$ respectively. From the table, we see clearly that (1) our method consistently, sometimes significantly, yields more accurate clustering than spectral relaxation approach. In particular, the results are significantly improved when $z_{\rm out} = 8$ (Table \ref{tab1}); (2) both of these algorithms achieved best performance when the parameter $\lambda = 0.4$. This is partly due to the fact that the artificial communities are of equal size and similar total degree \cite{Li08}. In fact, how to choose an appropriate value for parameter $\lambda$ seems not a trivial issue. We omit it here due to out of the scope of this paper and suggest interested readers seeing \cite{Li08} for more discussions.

\subsubsection{The LFR benchmarks} \label{sec5:sub1:sub2}

\begin{figure}
\centering
\includegraphics[width=0.8\textwidth]{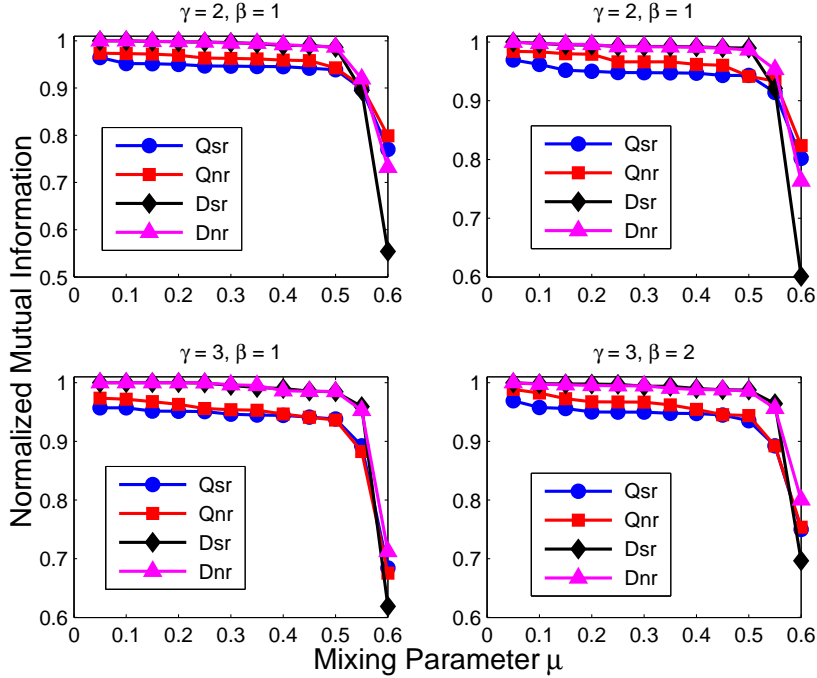}
\caption{Test of our algorithm on the LFR benchmark graphs \cite{Lan08,Lan09} where the number of nodes $n = 500$ and average degree $\langle d \rangle = 15$. Each point corresponds to an average over $50$ realizations. The term $Q_{sr}$ denotes maximizing the $Q$-measure based on the spectral relaxation approach and similar for the other three terms. For the normalized mutual information, our method always outperforms the spectral relaxation approach, especially when $\mu$ is larger than the threshold $\mu_c = 0.5$ beyond which communities are no longer defined in the strong sense.}
\label{fig2}
\end{figure}

The LFR benchmark \cite{Lan08,Lan09} is a realistic graphs for community detection, which takes the heterogeneity of both node degree and community size into account. The node degree and community size distributions are both power law, with exponents $\gamma$ and $\beta$, respectively. The number of nodes is $n$, and the average degree is $\langle k \rangle$. In the construction of the benchmark networks, each node receives
its degree once and for all and keeps it fixed until the end. It is more practical to choose as independent parameter the mixing parameter $\mu$, which expresses the ratio between the external degree of a node with respect to its community and
the total degree of the node. To compare the built-in modular structure with the one delivered by the algorithm, we adopt
here the normalized mutual information (NMI), which has proved to be reliable \cite{Lan08,Lan09}.

\begin{figure}
\centering
\includegraphics[width=0.8\textwidth]{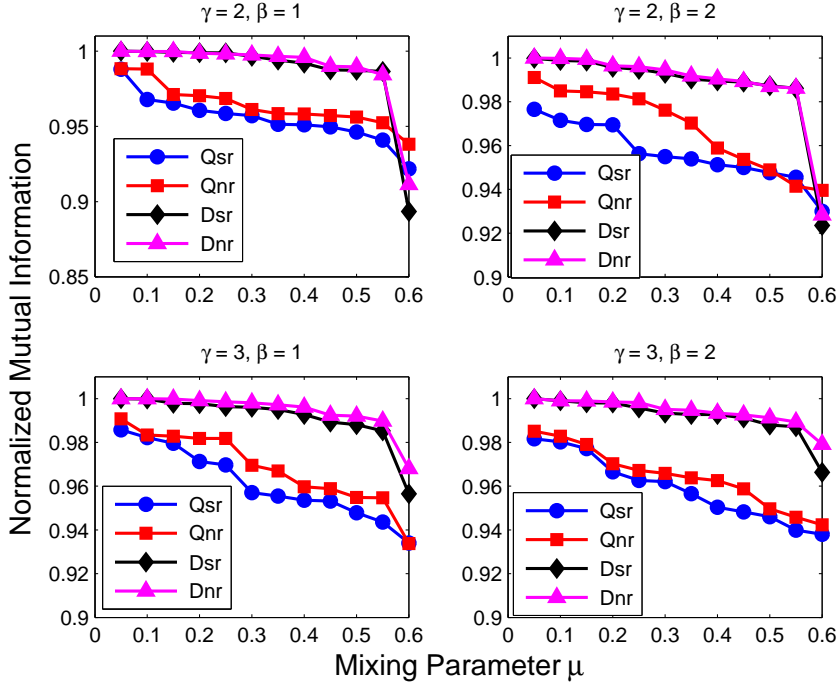}
\caption{Test of our algorithm on the LFR benchmark graphs \cite{Lan08,Lan09} where the number of nodes $n = 500$ and average degree $\langle d \rangle = 20$. Each point corresponds to an average over $50$ realizations. For the normalized mutual information, our method always outperforms the spectral relaxation approach, especially when $\mu$ is larger than the threshold $\mu_c = 0.5$ beyond which communities are no longer defined in the strong sense.}
\label{fig3}
\end{figure}

\begin{figure}
\centering
\includegraphics[width=0.8\textwidth]{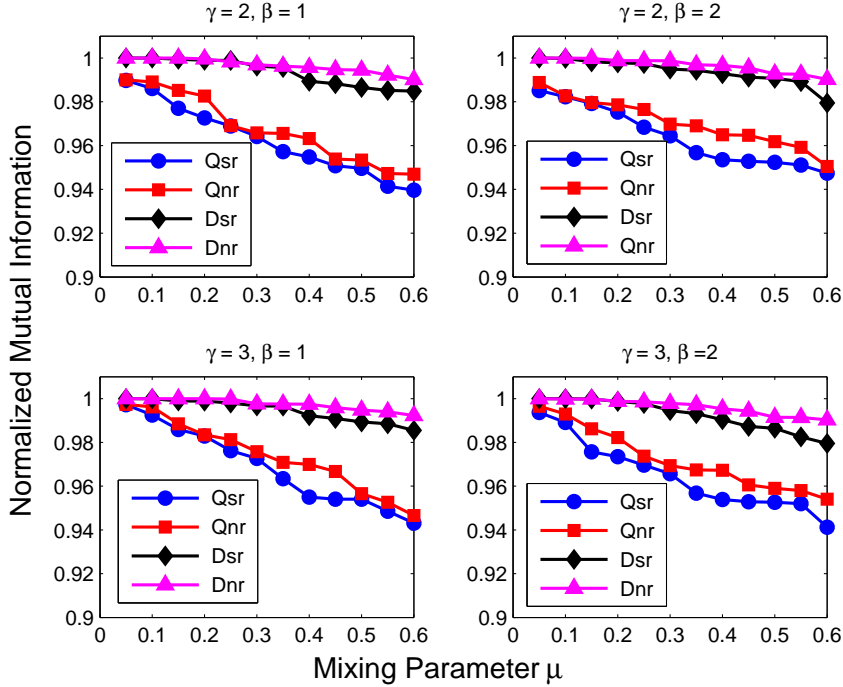}
\caption{Test of our algorithm on the LFR benchmark graphs \cite{Lan08,Lan09} where the number of nodes $n = 500$ and average degree $\langle d \rangle = 25$. Each point corresponds to an average over $50$ realizations. For the normalized mutual information, our method always outperforms the spectral relaxation approach, especially when $\mu$ is larger than the threshold $\mu_c = 0.5$ beyond which communities are no longer defined in the strong sense.}
\label{fig4}
\end{figure}

In Fig. \ref{fig2}-\ref{fig4}, we show the results when we apply our algorithm to the LFR benchmark problem with $n = 500$ and $\langle d \rangle = 15, 20, 25$. The four panels correspond to the results with the four pairs of exponents $(\gamma, \beta) = (2,1), (2,2), (3,1), (3,2)$, respectively. We have chosen combinations of the extremes of the exponent ranges in order to explore the widest spectrum of network structures. Each curve shows the variation of the normalized mutual information with the mixing parameter $\mu$. Clearly, the results depend on all parameters of the benchmark, from the exponents $\gamma$ and $\beta$ to the average degree $\langle d \rangle$. We can see that (1) the performance of our method is better when the average degree $\langle d \rangle$ becomes large, whereas it gets worse when the mixing parameter $\mu$ becomes larger; (2) For the normalized mutual information, our results are all above $0.9$ when $\mu$ is less than the threshold $\mu_c = 0.5$ that marks the the border beyond which communities are no longer defined in the strong sense, i.e., such that each node has more neighbors in its own community than in the others; (3) whatever modularity function was used, our method always achieves the superior performance on all the LFR benchmark problems, especially for the most difficult case when the mixing parameter $\mu$ is large than $\mu_c$. These observations support the effectiveness of our algorithm.

\subsection{Applications to real world networks} \label{sec5:sub2}

\subsubsection{Zachary's karate-club network} \label{sec5:sub2:sub1}

This famous network was constructed by Zachary after he observed social interactions between 34
members of a karate club at an American University for a period of two years \cite{Zac77}. Soon after, a
dispute arose between the club¡¯s administrator and instructor and thus the club split into two smaller
ones. The question concerned is if we can uncover the potential behavior of the network, detect the
two communities, and in particular, identify which community a node belongs to. It has become a
prototype to test the algorithms for finding community structure in networks \cite{New04,Li08,Jia09}.

\begin{figure}
\centering
\includegraphics[width=0.5\textwidth]{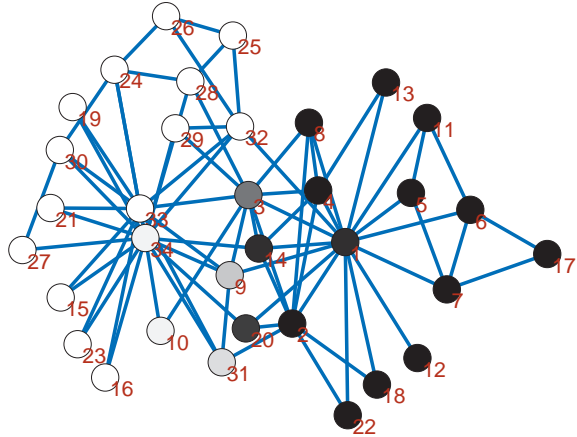}
\caption{The partition of karate club network obtained by our new method based on $D$-measure. The gray-scale plot of $p_1$ and $p_2$ for each node. The darker the color, the larger the value of $p_1$. The transition nodes or intermediate nodes are clearly shown.}
\label{fig5}
\end{figure}

We apply here our new method to this network based on $D$-measure. In \cite{Zac77}, Zachary gave the partition $\mathbb{S}_1 = \{1:8,11:14,17,18,20,22\}$ and $\mathbb{S}_2 = \{9,10,15,16,19,21,23:34\}$. Thus, we first employ our method as a hard clustering algorithm (Eq.(\ref{eq19}) in Section \ref{sec3}) and obtain Acc = 1.000 and NMI = 1.000 which is slightly better than that obtained by spectral relaxation approach Acc = 0.9706 and NMI = 0.8361. Fig.\ref{fig5} gives the partition that divides the network into two groups of roughly equal size and produces a completely consistent split with the actual division of the original club. We color each node as black or white in the figure to show its attribute in the graph representation. From the viewpoint of the soft clustering, the attribute of each node is no longer an indicator function but rather a discrete probability distribution. In our following notations,
the association probability $p_1$ or $p_2$ indicates the probability
of each node belonging to black or white colored group,
respectively.

\begin{table}
\caption{\label{tab2}The association probability and community entropy of each node. $p_1$ and $p_2$ are the probabilities of belonging to the black or white colored groups in Fig.\ref{fig5}, respectively. Larger entropies mean that the corresponding nodes are more difficult to classify
into communities.}
\centering
\scalebox{0.8}{
\begin{tabular}{c c c c c c c c c c c c c c }
\hline
\hline
  Nodes & 1 & 2 & 3 & 4 & 5 & 6 & 7 & 8 & 9 & 10 & 11 & 12\\
  $p_1$ & 0.9669& 1.000 & 0.6458 & 1.000 & 1.000 & 1.000 & 1.000 & 1.000 & 0.2726 &	0.0689 & 1.000 & 1.000\\
  $p_2$ & 0.0331& 0 &	0.3542 & 0 &	0	&0	&0	& 0&	0.7274 &	0.9311&	0&	0\\
  $\xi_i$ & 0.1454& 0 &	0.6500 &0 &	0&	0&	0&	0&	0.5859 &	0.2507 &	0&	0\\
  \hline
  Nodes & 13 & 14 & 15 & 16 & 17 & 18 & 19 & 20 & 21 & 22 & 23 & 24\\
  $p_1$ & 1	& 0.9373&0 &0&1.000	&1.000 &	0	& 0.8936 &	0	&1.000&	0&	0\\
  $p_2$ & 0	& 0.0627 &	1.000&	1.000&	0&	0&	1.000&	0.1065 &	1.000&	0&	1.000&	1.000\\
  $\xi_i$ & 0 & 0.2343&	0	&0&	0	&0&	0	&0.3392 &	0	&0	&0&	0\\
  \hline
  Nodes & 25 & 26 & 27 & 28 & 29 & 30 & 31 & 32 & 33 & 34 &  & \\
  $p_1$ & 0	& 0	 & 0 &	0 &	0 &	0 & 0.1406 & 0.0063 &	0 &	0.0324 & & \\
  $p_2$ & 1.000 & 1.000&1.000&	1.000&	1.000&	1.000&	0.8594&	0.9937&	1.000&	0.9676 & &\\
  $\xi_i$ & 0 & 0 & 0 & 0 & 0 &	0 & 0.4061 & 0.0380 &	0 &	0.1431 & & \\
  \hline
  \hline
\end{tabular}}
\end{table}

We summarized the association probability and community entropy of each node in Table \ref{tab2}. From Table \ref{tab2}, we find $p_1 = 1$ for nodes $\{2,4:8, 11:13, 17,18,22\}$, which lie at the boundary of the black colored group (Fig.\ref{fig5}), and $p_2 = 1$ for nodes $\{15,16,19,21,23:30,33\}$, which mostly lie at the boundary of the white colored group (except node 33 that lies at the center of the white colored group). The nodes $\{3, 9, 10, 14, 20, 31\}$ have more diffusive probability and higher community entropy value (Table \ref{tab2}) and they play the role of transition nodes between the black and white colored groups. In other words, these nodes constitute the fuzzy boundaries of two communities (Fig.\ref{fig5}). In particular, node 3 is the most diffusive node of those that are exactly in-between the two smaller clubs. This means that such members have good friendship with the two clubs at the same time. We
can visualize the data $p$ more transparently with the gray-scale
plot for each node shown in Fig.\ref{fig5}. Also we can uncover
more such nodes with a different degree of instability according
to the sorted $\xi$ values.

\subsubsection{Journal citation network}\label{sec5:sub2:sub2}

\begin{figure}
  \centering
  \subfigure[]{
    \label{fig6:suba}
    \includegraphics[width=0.4\textwidth]{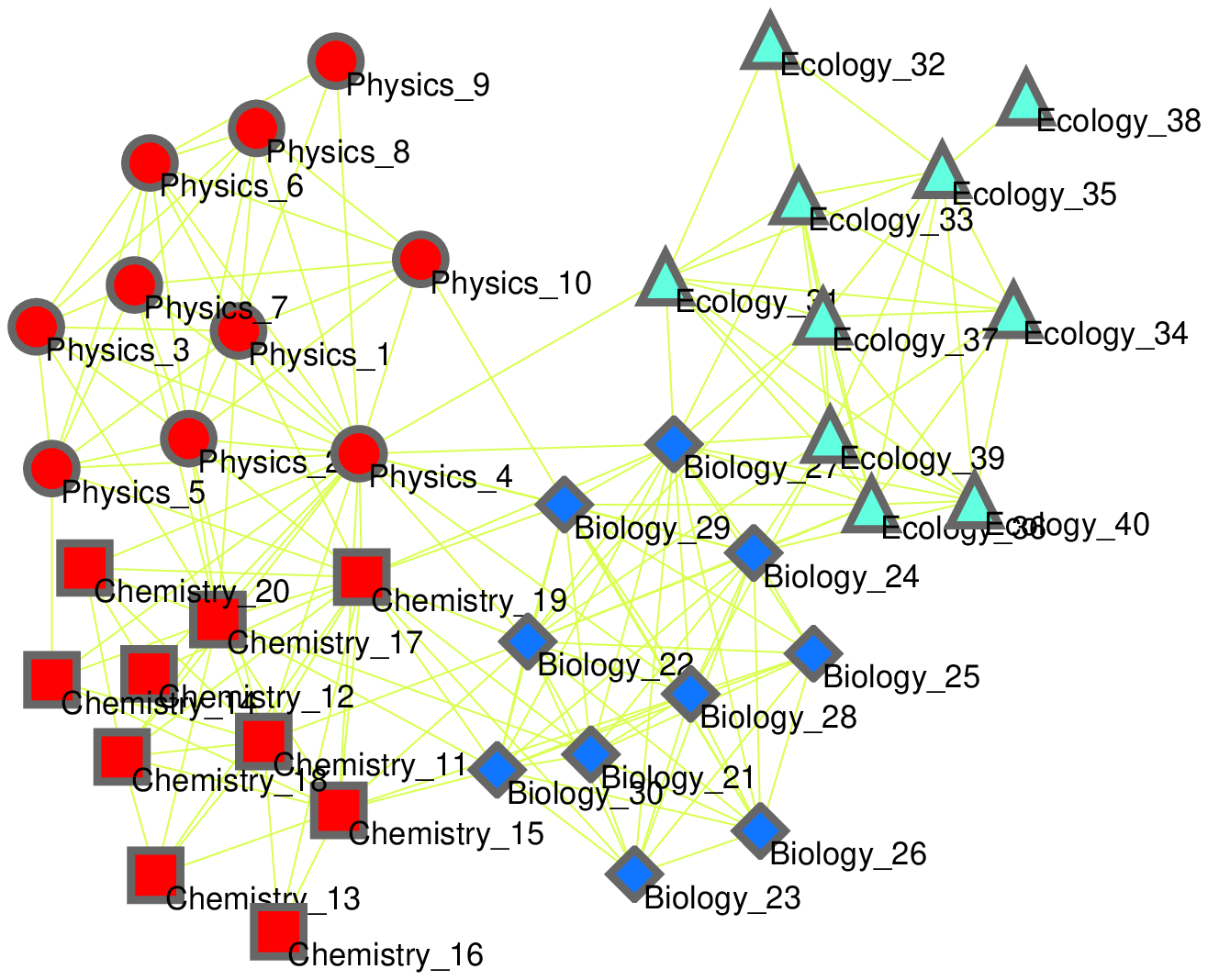}}
  \subfigure[]{
    \label{fig6:subb}
    \includegraphics[width=0.4\textwidth]{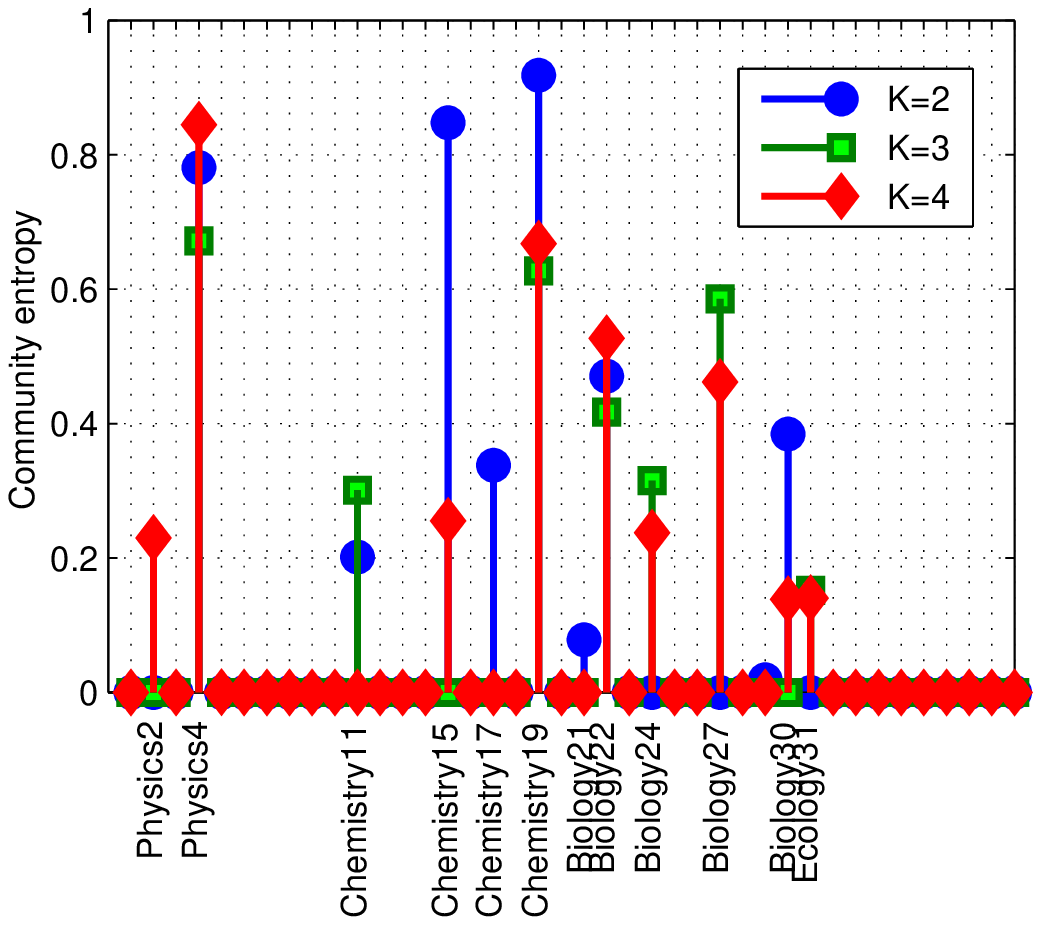}}
  \caption{(a) Community structure of the journal citation network extracted by our algorithm. The ellipse, squares, diamonds and triangles denote the split of the network into four groups that are perfectly consistent with the original partition. The different colours denote the partition of the network into two clusters, and the blue nodes are further subdivided into two smaller groups denoted by the different shades of vertices. (b) Soft clustering of the journal citation network. The blue, green and red bars illustrate the association probability of each node when the network is partitioned into two, three, and four modules, respectively.}
  \label{fig6}
\end{figure}

The journal citation network, proposed in \cite{Ros07}, consists of 40 journals as nodes from 4 different fields: physics, chemistry, biology and ecology, and 189 links connecting nodes if at least one article from one journal cites an article in the other journal during 2004. Ten
journals with the highest impact factor in the 4 different
fields were selected. With the $D$-measure at hand, we partition the network into 4 communities as that obtained by spectral relaxation approach (see Fig.\ref{fig6:suba}). This split of the network is consistent perfectly with its orignial partition, i.e., Acc = 1 and NMI = 1. The community entropy of each node is given by the red bar in Fig.\ref{fig6:subb} where the nodes $\{\mathrm{Physics\_4}, \mathrm{Chemistry\_19}, \mathrm{Biology\_22}, \mathrm{Biology\_27}\}$ form the fuzzy boundary of these four communities. In fact, these nodes are densely linked to their neighbors in different communities simultaneously. Particularly, the node $\mathrm{Physics\_4}$ (Physical Review Letters), which was misclassified into chemistry cluster in \cite{Ros07} and is correctly classified here, has 9 links to physics, 8 links to chemistry, 5 links to biology and only 1 link to ecology.

We also partition the network into two, three, or five modules as suggested by the previous studies \cite{Li08,Ros07}. When we split the network into two components, physical journals group together with chemical journals, and biological journals cluster together with ecological journals. In this scenario, nodes $\{\mathrm{Physics\_4}, \mathrm{Chemistry\_15}, \mathrm{Chemistry\_19}\}$ become the transition ones between these two communities (the blue bar in Fig.\ref{fig6:subb}). When we partition it into three clusters, ecological journals and biological journals separate, but physical journals and chemical journals remain together in a single module. At this time, nodes $\{\mathrm{Physics\_4}, \mathrm{Chemistry\_19}, \mathrm{Biology\_22}, \mathrm{Biology\_27}\}$ are the unstable nodes which has the largest entropy values (the green bar in Fig.\ref{fig6:subb}). When we further split the network into five communities, the partition is essentially the same as with four, only with the singly connected journal $\mathrm{Ecology\_38}$ (Conservation Biology) split off by itself as a community. All the results are consistent with that in \cite{Li08,Ros07}.

\subsubsection{American college football team network}\label{sec5:sub2:sub3}

The third real network is the college football team network of the United States. The schedule of Division I games can be represented by a network, in which the nodes denote the 115 teams and the edges represent 613 games played in the course of the year. These teams are divided into 12 conferences containing around 8-12 teams each. Games are more frequent between members of the same conference than they are between members of different conferences, with teams playing an average of about seven intra-conference games and four inter-conference games in the 2000 season. Inter-conference
play is not uniformly distributed; teams that are geographically
close to one another but belong to different conferences
are more likely to play with one another than teams
separated by large geographic distances. The natural community
structure in the network makes it a commonly used
workbench for community-detecting algorithm testing \cite{Gir02,New04,Li08}.

\begin{figure}
\centering
\centering
  \subfigure[]{
    \label{fig7:suba}
    \includegraphics[width=0.45\textwidth]{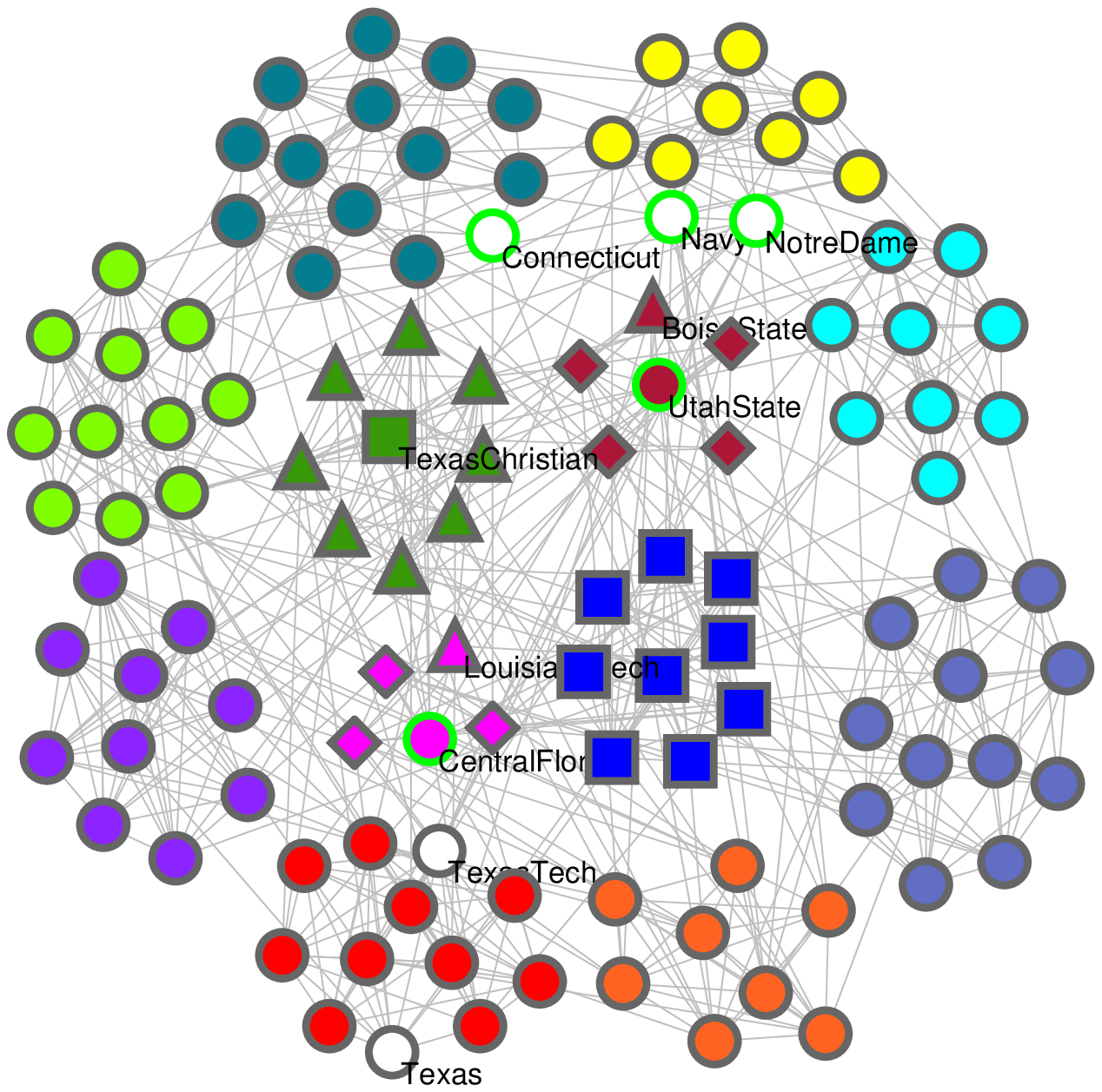}}
  \subfigure[]{
    \label{fig7:subb}
    \includegraphics[width=0.45\textwidth]{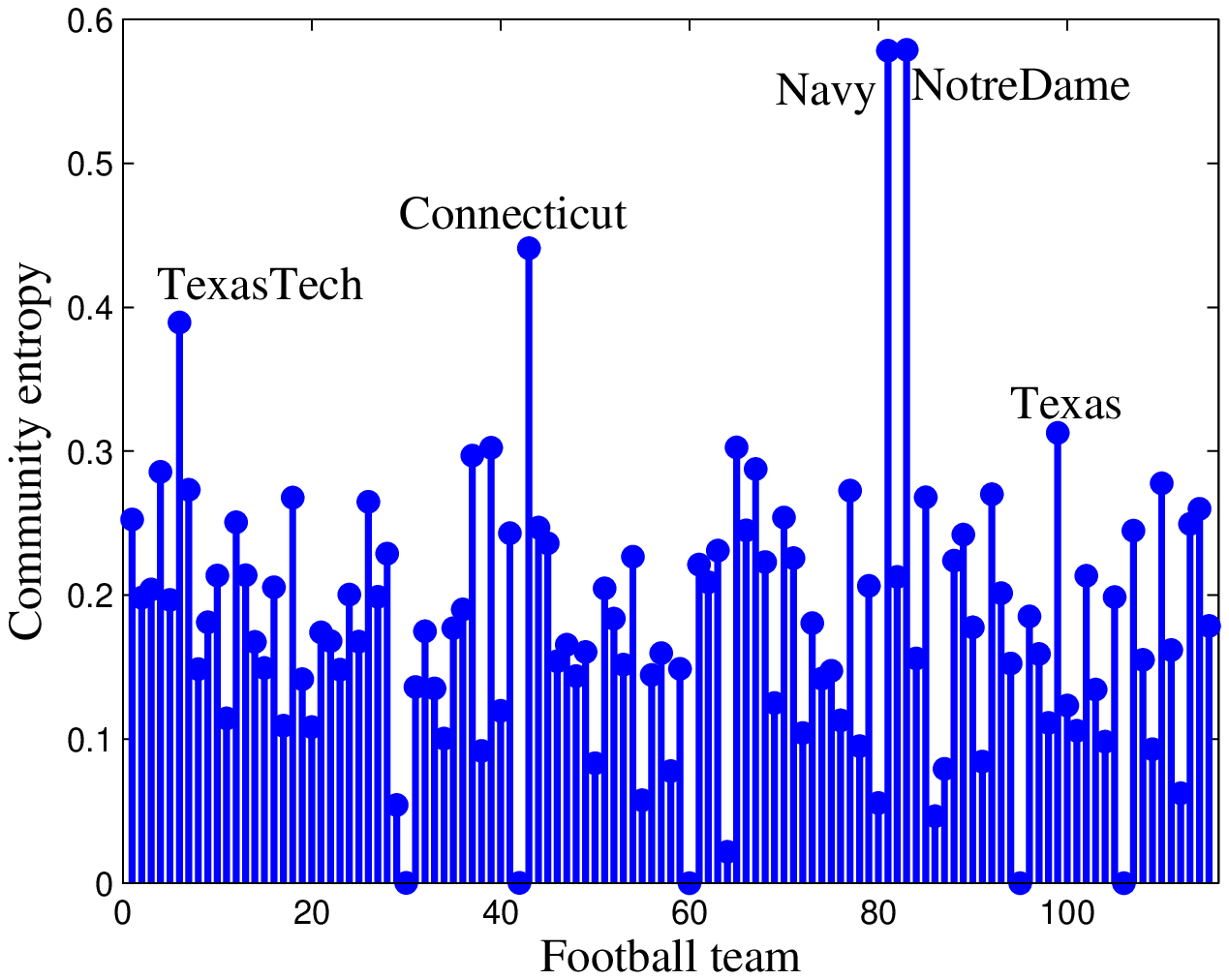}}
\caption{(a) Community structures of American college football team network. Nodes are colored according to their different conferences. The 5 white colored nodes have the highest community entropy values and are the most difficult to classify into communities. (b) The community entropy of each node. The larger value means that corresponding node is more difficult
to classify into communities.}
\label{fig7}
\end{figure}

Using our algorithm, we can split the network into conferences with a high degree of success,i.e., Acc = 0.9130 and NMI = 0.9232. Fig.\ref{fig7:suba} shows the community structure of football team network calculated by our method. We further investigate the 10 misclassified teams. 4 teams $\{$\lq \lq Connecticut \rq \rq, \lq \lq Navy \rq \rq, \lq \lq NotreDame \rq \rq, \lq \lq UtahState \rq \rq $\}$ belong to the conference \lq \lq Independents\rq \rq (denoted as green-edge nodes) which means each member of this conference is not associated with any existing football conference and thus there are few edges between its 5 members distributed to other communities, but they tend to be clustered with the conference which they are most closely associated with. Specifically, \lq \lq Connecticut\rq \rq is distributed to community \lq \lq Med-American\rq \rq, \lq \lq Navy \rq \rq and \lq \lq NotreDame \rq \rq are classified into community \lq \lq Big East \rq \rq, and \lq \lq UtahState \rq \rq is assigned to community \lq \lq Sun Belt\rq \rq. The remaining 6 teams, $\{$ \lq \lq LouisianaMonroe \rq \rq, \lq \lq MiddleTennesseeState \rq \rq, \lq \lq LouisianaLafayette \rq \rq$\}$ associated with conference \lq \lq Sun Belt\rq \rq (diamond nodes) and $\{$ \lq \lq BoiseState \rq \rq, \lq \lq Houston \rq \rq$\}$ associated with conference \lq \lq Western Athletic\rq \rq, are misclassified as \lq \lq Independent \rq \rq. This happens partly because the \lq \lq Sunbelt\rq \rq teams played nearly as many games against
\lq \lq Western Athletic \rq \rq teams (triangle nodes) as they played in their own conference, and they also played quite a number of games against \lq \lq Mid-American\rq\rq teams (square nodes) \cite{Gir02,Li08}. All these teams are incorrectly classified due to the fact that there are more games with the teams in the classified communities than there are with the teams in their
own conferences. The community entropy of each team are shown in Fig.\ref{fig7:subb}. The nodes with top 5 highest entropy value are teams $\{$ \lq \lq Navy \rq \rq, \lq \lq NotreDame \rq\rq, \lq \lq Connecticut\rq \rq,\lq \lq TexasTech \rq \rq, \lq \lq Texas \rq \rq$\}$ (see the white colored nodes in Fig.\ref{fig7:subb}). Clearly, these teams are both play enough games with other ones belonging to various different conference. For example, the team \lq \lq Navy\rq \rq has with the highest entropy value, and it play game with other teams belonging to as many as 6 different conferences, i.e., \lq \lq Big East\rq \rq (three times), \lq \lq Atlantic Coast\rq \rq (twice), \lq \lq Conference USA \rq \rq(twice), \lq \lq Mid-American \rq \rq (once), \lq \lq Mountain West\rq \rq(once) and \lq \lq Western Athletic \rq \rq(once). All these results suggest that the community structure found by our
method seems to reveal a more precise organization than
the original conferences.

\section{Concluding remarks}\label{sec6}

In this paper, we first show that the problem of maximizing one-parameter $\lambda$ family of quantitative functions, encompassing both the modularity ($Q$-measure) and modularity density ($D$-mesure), for community detection can be understood as a combinatoric optimization involving the trace of a matrix called modularity-Laplacian.  After that, we proposed the nonnegative relaxation to solve such optimization problem, and introduced efficient algorithms to solve this problem with explicit nonnegative constraint rigorously. Differing from the standard spectral relaxation approach, our solutions are very close to the ideal class indicator matrix and can directly assign nodes into communities. In addition, we prove the convergence and correctness of our proposed method. Extensive experimental results on Newman artificial networks and LFR benchmark networks show that the proposed algorithms always outperform, sometimes significantly, when compared with the traditional spectral clustering method. Furthermore, We emphasize the soft clustering nature of this method due to its inherent near-orthogonality of columns which can be reformulated as the posterior probability of corresponding node belonging to each community. Simulation results on real world networks illustrate that our algorithm can be applied to fuzzy community detection and unstable node discovery, which can help to elucidate a network's intrinsic structure and facilitate our analysis.

\section*{Acknowledgments}
The authors thank Mu Li for helping to generate the \textit{ad hoc} artificial networks, Professor M.E.J. Newman for providing data of karate club network and American college football team network and useful comments, and Martin Rosvall for providing data of journal index network.

\appendix
\section{Proof of Theorem \ref{the2}} \label{app}

We use the auxiliary function approach \cite{Lee01}. An auxiliary function $G(H,H)$ of function $L(H)$ satisfies $G(H,H) = L(H)$, $G(H,\tilde{H}) \le L(H)$. We define

\begin{equation}
H^{(t+1)} = \arg \max_H G(H,H^{(t)})
\label{eq16}
\end{equation}
Then by construction, we have
\begin{equation}
L(H^{(t)}) = Z(H^{(t)}, H^{(t)}) \le Z(H^{(t+1)}, H^{(t)}) \le L(H^{(t+1)})
\label{eq17}
\end{equation}
This proves that $L(H^{(t)})$ is monotonically increasing. We write Eq.(\ref{eq15}) as
$$
L = {\rm Tr}[Q^T (\rho I + M^-)Q + \Lambda^- Q^TQ - Q^T M^+ Q - \Lambda^+ Q^TQ]
$$
We can show that one auxiliary function of $L$ is
\begin{eqnarray}
Z(H,\tilde{H}) &=& \sum_{ijk} (\rho + M^-)_{ij} \tilde{H}_{ik} \tilde{H}_{jk}(1 + \log\frac{H_{ik}H_{jk}}{\tilde{H}_{ik}\tilde{H}_{jk}}) \nonumber\\
& & + \sum_{ilk}(\Lambda^-)_{kl}\tilde{H}_{ik}\tilde{H}_{il}(1 + \log\frac{H_{ik}H_{il}}{\tilde{H}_{ik}\tilde{H}_{il}}) \nonumber\\
& & - \sum_{ik} \frac{(M^+\tilde{H})_{ik}H^2_{ik}}{\tilde{H}_{ik}}- \sum_{ik} \frac{(\tilde{H}\Lambda^+)_{ik}H^2_{ik}}{\tilde{H}_{ik}} \nonumber
\end{eqnarray}
using the inequality
$$
z \ge 1 + \log z,~~~z = H_{ik}H_{jk}/\tilde{H}_{ik}\tilde{H}_{jk}
$$
and a generic inequality
$$
\sum_{i=1}^n \sum_{p=1}^k \frac{(AS'B)_{ip}S^2_{ip}}{S'_{ip}} \ge {\rm Tr}(S^TASB)
$$
where $A,B,S,S' > 0$, $A = A^T$, $B = B^T$. We now find the global maxima of $Z(H) = G(H,\tilde{H})$. The gradient is
\begin{eqnarray}
\frac{\partial Z(H,\tilde{H})}{\partial H_{ik}}&=& 2\frac{[(M^- + \rho)\tilde{H}]_{ij}\tilde{H}_{ik}}{H_{ik}}+ 2\frac{(\tilde{H}\Lambda^-)_{kl}\tilde{H}_{ik}}{H_{il}} \nonumber\\
& & - 2\frac{(M^+\tilde{H})_{ik}\tilde{H}_{ik}}{H_{ik}}-2\frac{(\tilde{H}\Lambda^+)_{ik}\tilde{H}_{ik}}{H_{ik}}
\end{eqnarray}
The second derivative
\begin{eqnarray}
\frac{\partial^2 G(H,\tilde{H})}{\partial H_{ik}\partial H_{jl}} & = & -2W_{ik} \delta_{ij} \delta_{kl} W_{ik} \nonumber\\
& = & \frac{[(M^- + \rho)\tilde{H}]_{ij}\tilde{H}_{ik}}{H^2_{ik}}+ \frac{(\tilde{H}\Lambda^-)_{ik}\tilde{H}_{ik}}{H^2_{il}} \nonumber\\
& & +\frac{(M^+\tilde{H})_{ik}}{H_{ik}}+\frac{(\tilde{H}\Lambda^+)_{ik}}{H_{ik}}
\end{eqnarray}
is negative definite. Thus $Z(H)$ is a concave function in $H$ and has a unique global maximum. This maximum is obtained by setting the first derivative to zero, yielding:
\begin{equation}
H^2_{ik} = H^2_{ik} \frac{[(M^- + \rho)\tilde{H}]_{ij} + (\tilde{H}\Lambda^-)_{ik}}{(M^+\tilde{H})_{ik} + (\tilde{H}\Lambda^+)_{ik}}
\label{eq17}
\end{equation}
According to Eq (\ref{eq16}), $H^{(t+1)} = H$ and $H^{(t)} = \tilde{H}$, we see that Eq (\ref{eq17}) is the update rule of Eq (\ref{eq7}). Thus Eq (\ref{eq17}) always holds.


\begin{thebibliography}{99}

\bibitem{Bar02}
A. Barab\'{a}si, The New Science of Networks, Perseus Publishing,
Cambridge, MA, 2002.

\bibitem{Ding05}
C.H.Q. Ding, X. He, H. Simon, On the equivalence of nonnegative matrix
factorization and spectral clustering, in Proc. SIAM. Data Mining Conference, 2005, pp. 606--610.

\bibitem{For07}
S.Fortunato, M.Barth\'{e}lemy, Resolution limit in community detection,
Proc. Natl. Acad. Sci. USA 104 (2007) 36--41.

\bibitem{For10}
S. Fortunato, Community detection in graphs, Physics Reports 486 (2010) 75--174.

\bibitem{Gfe05}
D. Gfeller, J.C. Chappelier, P.D.L. Rios, Finding instabilities in the community
structure of complex networks, Phys Rev E 72 (2005) 056135.

\bibitem{Gir02}
M. Girvan, M. E. J. Newman, Community structure in social and biological
networks, Proc. Natl. Acad. Sci. USA 99 (2002) 7821--7826.

\bibitem{Jia09}
J.Q. Jiang, A.W.M. Dress, G. Yang, A spectral clustering-based framework for detecting community structures in complex networks, Appl. Math. Lett. 22 (2009) 1479--1482.

\bibitem{Lan08}
A. Lancichinetti, S. Fortunato, F. Radicchi, Benchmark graphs for testing community
detection algorithms. Phys. Rev. E, 78 (2008) 046110.

\bibitem{Lan09}
A. Lancichinetti, S. Fortunato, Community detection algorithms: a comparative analysis,
Phys. Rev. E, 80 (2009) 056117.

\bibitem{Lee01}
D.D. Lee, H.S. Seung, Algorithms for Non-negative Matrix Factorization, Advances in Neural Information Processing Systems 13: Proceedings of the 2000 Conference. MIT Press. pp. 556--562.

\bibitem{Li08}
Z. Li, S. Zhang, R. S. Wang, X. S. Zhang, L. Chen, Quantitative
function for community detection, Phys. Rev. E, 77 (2008) 036109.

\bibitem{Lov86}
L. Lovasz, M. Plummer, Matching Theory, Akademiai Kiado, North Holland, Budapest, 1986.

\bibitem{Luo09}
D. Luo, C.H.Q. Ding, H. Huang, T. Li, Non-negative Laplacian Embedding,  in Proc. International Conference on Data Mining, 2009, pp.337--346.


\bibitem{New01}
M.E.J. Newman, Scientific collaboration networks: I. Network construction and fundamental
results, Phys. Rev. E 64 (2001) 016131.

\bibitem{New04}
M.E.J. Newman, M.Girvan, Finding and evaluating community structure in networks, Physical Review E 69 (2004) 026113.

\bibitem{Pal05}
G. Palla, I. Der\'{e}nyi, I. Farkas, T. Vicsek, Uncovering the overlapping
community structure of complex networks in nature and society, nature 435 (2005) 814--818.

\bibitem{Por09}
M. Porter, J.P. Onnela,P. Mucha, Communities in networks, Notices of the American Mathematical Society, 56 (2009) 1082--1097.

\bibitem{Rei04}
J. Reichardt, S. Bornholdt, Detecting fuzzy communities in complex networks
with a Potts model, Phys. Rev. Lett. 93 (2004) 218701.

\bibitem{Ros07}
M. Rosvall, C.T. Bergstrom, An information-theoretic framework for resolving community, Proc. Natl. Acad. Sci. USA 104 (2007) 7327--7331.

\bibitem{Whi05}
S. White, P. Smyth,  A spectral clustering approach to finding communities in
graph, in: H. Kargupta, J. Srivastava, C. Kamath, A. Goodman (Eds.), Proceedings of the 5th SIAM International
Conference on Data Mining, Society for Industrial and Applied Mathematics, Philadelphia, 2005, pp. 274--285.

\bibitem{Zac77}
W.W. Zachary, An information flow model for conflict and fission in small
groups, J. Anthropol. Res. 33 (1977) 452--473.


\end{thebibliography}
\end{document}